\documentclass[12pt,preprint]{aastex}

\def\ltsima{$\; \buildrel < \over \sim \;$}
\def\gtsima{$\; \buildrel > \over \sim \;$}
\def\lsim{\lower.5ex\hbox{\ltsima}}
\def\gsim{\lower.5ex\hbox{\gtsima}}
\def\lapp{\ifmmode\stackrel{<}{_{\sim}}\else$\stackrel{<}{_{\sim}}$\fi}
\def\gapp{\ifmmode\stackrel{>}{_{\sim}}\else$\stackrel{<}{_{\sim}}$\fi}

\newdimen\minuswidth    
\setbox0=\hbox{$-$}
\minuswidth=\wd0
\catcode`@=\active
\def@{\kern\minuswidth}
\setbox0=\hbox{\rm0}
 
\shorttitle{The optical companion to the MSP M28H}
\shortauthors{Pallanca et al.}
 
\begin{document} 
\title{The optical companion to the binary millisecond pulsar
  J1824-2452H in the globular cluster M28\footnote{Based on
    observations with the NASA/ESA HST, obtained at the Space
    Telescope Science Institute, which is operated by AURA, Inc.,
    under NASA contract NAS5-26555. Also based on observations
    collected at the European Southern Observatory at La Silla (Chile)
    with the MPI Telescope under program 164.O-0561(F)}}

\author{
C. Pallanca\altaffilmark{2},
E. Dalessandro\altaffilmark{2},
F.R. Ferraro\altaffilmark{2},
B. Lanzoni\altaffilmark{2},
R.T. Rood\altaffilmark{3},
A. Possenti\altaffilmark{4},
N. D'Amico\altaffilmark{4,5},
P.C. Freire\altaffilmark{6},
I. Stairs\altaffilmark{7},
S.M. Ransom\altaffilmark{8},
S. B\'egin\altaffilmark{7,9},
}
\affil{\altaffilmark{2} Dipartimento di Astronomia, Universit\`a degli Studi
di Bologna, via Ranzani 1, I--40127 Bologna, Italy}

\affil{\altaffilmark{3} Astronomy Department, University of Virginia,
  P.O. Box 3818, Charlottesville, VA 22903-0818, USA}

\affil{\altaffilmark{4} INAF-Osservatorio Astronomico di Cagliari,
  localit\`a Poggio dei Pini, strada 54, I-09012 Capoterra, Italy}

\affil{\altaffilmark{5} Dipartimento di Fisica, Universit\`a di Cagliari, 
Cittadella Universitaria, I-09042 Monserrato, Italy}

\affil{\altaffilmark{6} Max-Planck-Institut f$\ddot{u}$r
  Radioastronomie, Auf dem H$\ddot{u}$gel 69, D-53121 Bonn, Germany}

\affil{\altaffilmark{7} Department of Physics and Astronomy,
  University of British Columbia, 6224 Agricultural Road, Vancouver,
  BC, V6T 1Z1, Canada}

\affil{\altaffilmark{8} National Radio Astronomy Observatory,
  Charlottesville, VA 22903, USA}

\affil{\altaffilmark{9} D\'epartement de physique, de g\'enie physique
  et d'optique, Universit\'e Laval, Qu\'ebec, QC, G1K 7P4, Canada}

\date{8 October, 2010}

\begin{abstract}
We report on the optical identification of the companion star to the
eclipsing millisecond pulsar PSR J1824-2452H in the galactic globular
cluster M28 (NGC 6626). This star is at only $0.2\arcsec$ from the
nominal position of the pulsar and it shows optical variability ($\sim
0.25$ mag) that nicely correlates with the pulsar orbital period.  It
is located on the blue side of the cluster main sequence, $\sim 1.5$
mag fainter than the turn-off point. The observed light curve shows
two distinct and asymmetric minima, suggesting that the companion star
is suffering tidal distortion from the pulsar. This discovery
increases the number of non-degenerate MSP companions optically
identified so far in globular clusters (4 out of 7), suggesting that
these systems could be a common outcome of the pulsar recycling
process, at least in dense environments where they can be originated
by exchange interactions.
\end{abstract} 


\section{INTRODUCTION}
The ultra-dense cores of globular clusters (GCs) are very efficient
``furnaces" for generating exotic objects, such as low-mass X-ray
binaries, cataclysmic variables, millisecond pulsars (MSPs), and blue
stragglers (e.g., Bailyn 1995; Verbunt et al. 1997; Ferraro et
al. 2001a). Most of these objects are thought to be the result of the
evolution of various kinds of binary systems originated and/or
hardened by stellar interactions (e.g., Clark 1975; Hills \& Day 1976;
Bailyn 1992; Ferraro et al. 1995). The nature and even the existence
of binary by-products are strongly related to the cluster core
dynamics (e.g. Heggie 1975, Ivanova et al. 2008). Hence they may be
very useful diagnostics of the dynamical evolution of GCs (e.g.,
Goodman \& Hut 1989; Hut et al. 1992; Meylan \& Heggie 1997; Pooley et
al. 2003; Fregeau 2008; Ferraro et al. 2009a).

In particular, MSPs are formed in binary systems containing a slowly
rotating neutron star (NS) that is eventually spun up to millisecond
periods by heavy mass accretion from an evolving companion, that, in
turn, is expected to become a white dwarf (WD; e.g. Lyne et al. 1987;
Alpar et al. 1982; Bhattacharya \& van den Heuvel 1991).  More than
50\% of known MSPs are found in GCs, although the Galaxy mass is 100
times larger than that of the whole GC system.  While this is
partially due to the deeper investigations performed in GCs with
respect to the field, it very likely indicates that dynamics plays a
significant role in the formation of these objects. In fact, in the
ultra-dense GC cores, dynamical interactions can promote the formation
of binaries suitable for recycling NSs into MSPs (e.g., Davies \&
Hansen 1998), while in the Galactic field the only viable formation
channel for MSPs is the evolution of primordial binaries.  The optical
identification of the companion stars to binary MSPs is a fundamental
step for characterizing these systems and for clarifying the possible
recycling mechanisms. In the case of GCs, it also represents a crucial
tool for quantifying the occurrence of dynamical interactions,
understanding the effects of crowded stellar environments on the
evolution of binaries, determining the shape of the GC potential well,
and estimating the mass-to-light ratio in the GC cores (e.g., Phinney
1992; Possenti et al. 2003; Ferraro et al. 2003a).

Despite their importance, up to now only six optical counterparts to
MSP companions have been identified in five GCs.  In the
colour-magnitude diagram (CMD) three of them have positions consistent
with the cooling sequences of helium WDs, in agreement with the
expectations of the MSP recycling scenario. These are the companions
to MSP-U in 47 Tucanae (Edmonds et al. 2001); MSP-A in NGC 6752
(Ferraro et al. 2003b), and PSR B1620-26 in M4 (Sigurdsson et
al. 2003). The other identified companions show, instead, quite
peculiar properties.  The luminosity and colours of the optical
companion to MSP-A in NGC 6397 are totally incompatible with those of
a WD. This is a relatively bright, tidally deformed star, suggesting
that the system either harbours a newly born MSP, or is the result of
an exchange interaction (Ferraro et al. 2001b).  The companion star to
MSP-B in NGC 6266 is a similarly bright object, with luminosity
comparable to the cluster main sequence (MS) turn-off, an anomalous
red colour and optical variability suggestive of a tidally deformed
star which filled its Roche Lobe (Cocozza et al. 2008). This object is
also a Chandra X-ray source, thus supporting the hypothesis that some
interaction is occurring between the pulsar wind and the gas streaming
off the companion. Finally the companion to MSP-W in 47 Tuc has been
identified to be a faint MS star, showing large-amplitude, sinusoidal
luminosity variations probably due to the heating effect of the pulsar
(Edmonds et al. 2002).

As a part of a project aimed to perform a systematic search for
optical companions to binary MSPs in GCs (see Ferraro et al. 2001b,
Ferraro et al. 2003b, Cocozza et al. 2008), here we focus our
attention on M28 (NGC 6626). M28 is a Galactic GC with intermediate
central density ($\log \rho_0=4.9$ in units of $M_\odot/$pc$^3$; Pryor
\& Meylan 1993). It is the first GC where a MSP was discovered (Lyne
et al. 1987) and to date it is known to harbour a total of twelve
pulsars (B\'egin 2006). This is the third largest population of known
pulsars in GCs, after that of Terzan 5 (with 33 objects; Ransom et
al. 2005, but see the recent results by Ferraro et al. 2009b and
Lanzoni et al. 2010, suggesting that Terzan 5 is not a genuine GC) and
that of 47 Tuc (with 23 MSPs; Camilo et al. 2000; Freire et
al. 2003).\footnote{See the complete list of pulsars in GCs at
  http://www.naic.edu/$\sim$pfreire/GCpsr.html}

Among the binary MSPs harboured in M28, J1824-2452H (hereafter M28H)
deserves special attention since it is an eclipsing system showing a
number of timing irregularities, possibly due to the tidal effect on
the companion star (B\'egin 2006; Stairs et al. 2006).  It is located
at $\alpha_{2000}=18^{\rm h} 24^{\rm m} 31.61^{\rm s}$ and
$\delta_{2000}=-24^\circ 52' 17.2\arcsec$, it has an orbital period
$P=0.43502743$ days and shows eclipses for $\sim 20\%$ of it
(S. B\'egin et al. 2010, in preparation).  There is also an associated
X-ray source, possibly variable at the binary period and with a hard
spectrum (Bogdanov et al. 2010). Such X-ray emission is likely due to
the shock between the MSP magnetospheric radiation and the matter
released by the companion, like that detected in the case of MSP-W in
47 Tuc (Bogdanov et al. 2005). This further suggests that the
companion star to M28H is a non-degenerate star.  In this letter we
present its optical identification, based on high-quality,
phase-resolved photometry obtained with the new Wide Field Camera 3
(WFC3) on board the Hubble Space Telescope (HST).
 
\section{OBSERVATIONS AND  DATA ANALYSIS}
The photometric data-set used for this work consists of HST
high-resolution images obtained with the ultraviolet-visible (UVIS)
channel of the WFC3.  A set of supplementary HST Wide Field Planetary
Camera 2 (WFPC2) images, and ground-based wide-field images obtained
at the European Southern Observatory (ESO) have been retrieved from
the Science Archive and used for variability and astrometric purposes.

The WFC3 UVIS CCD consists of two twin detectors with a pixel-scale of
$\sim 0.04\arcsec$/pixel and a global field of view (FOV) of $\sim
162\arcsec\times 162\arcsec$.  The WFC3 images have been obtained on
2009 August 8 (Prop. 11615, P.I.  Ferraro) in four different bands.
The data-set consists of: 6 images obtained through the F390W filter
($\sim U$) with an exposure time of $t_{\rm exp}=800-850$ sec each; 7
images in F606W ($\sim V$) with $t_{\rm exp}=200$ sec; 7 images in
F814W ($\sim I$) with $t_{\rm exp}=200$ sec; and 7 images in F656N
(a narrow filter corresponding to H$\alpha$) with exposure time
ranging from $t_{\rm exp}=935$ sec, up to $t_{\rm exp}=1100$ sec.  All
the images are aligned and the cluster is almost centered in CHIP1.

Additional public WFPC2 images have been retrieved from the archive.
The first data-set (hereafter WFPC2-A) was obtained in 1997
(Prop. 6625) and consists of 8 images in F555W ($\sim V$) with $t_{\rm
  exp}=140$ sec and 9 images in F814W ($6\times t_{\rm exp}=160$ sec
and $3\times t_{\rm exp}=180$ sec).  The second sample (hereafter
WFPC2-B) consists of 13 images in F675W ($\sim R$), with $t_{\rm
  exp}=100$ sec each, secured in 2008 (Prop. 11340).

Finally, the wide-field data-set consists of 6 images in the $V$ and
$I$ filters, obtained in August 2000 with the Wide Field Imager (WFI)
at the ESO-MPI 2.2 m telescope (La Silla, Chile). The WFI consists of
a mosaic of eight chips, for a global FOV of $34'\times34'$.

The data reduction procedure (all the details will be discussed in a
forthcoming paper; E. Dalessandro et al. 2010, in preparation) has
been performed on the WFC3 ``flat fielded" (flt) images, once
corrected for ``Pixel-Area-Map" (PAM) by using standard IRAF
procedures. The photometric analysis has been carried out by using the
DAOPHOT package (Stetson 1987). For each image we modeled the point
spread function (PSF) by using a large number ($\sim 200$) of bright
and nearly isolated stars. Then we performed the PSF fitting by using
the DAOPHOT packages ALLSTAR and ALLFRAME (Stetson 1987, 1994). The
final star list consists of all the sources detected in at least 14
frames on a total number of 27. A similar procedure has been adopted
to reduce the WFPC2 images. For the WFPC2-A data-set we demanded that
sources were in at least 9 frames out of 17, whereas for the WFPC2-B
data-set in at least 7 frames out of 13. Since the WFC3 images heavily
suffer from geometric distortions within the FOV, we corrected the
instrumental positions of stars by applying the equations reported by
Bellini \& Bedin (2009) for the filter F336W, neglecting any possible
dependence on the wavelengths. Standard procedure (see e.g., Lanzoni
et al.  2007) has been adopted to analyze the WFI data.  Here we use
this data-set only for astrometric purposes.  In fact, very accurate
astrometry is the most critical task in searching for the optical
counterparts to MSPs, especially in crowded fields such as the central
regions of GCs, where primary astrometric standards are lacking (e.g.,
Ferraro et al. 2001a, 2003b).  For this reason we first placed the
wide-field catalog obtained from the WFI images on the absolute
astrometric system, and we then used the stars in common with the high
resolution data-sets as secondary standards.  In particular, the WFI
catalogue has been reported onto the coordinate system defined by the
Guide Star Catalogue II (GSCII) through cross-correlation. Then, we
placed the WFC3 and the WFPC2 catalogues on the same system through
cross-correlation with the WFI data-set.  Each transformation has been
performed by using several thousand stars in common, and at the end of
the procedure the typical accuracy of the astrometric solution was
$\sim0.2\arcsec$ in both right ascension ($\alpha$) and declination
($\delta$).

Finally, the WFC3 instrumental magnitudes have been calibrated to the
VEGAMAG system by using the photometric zero-points and the procedures
reported on the WFC3 web
page.\footnote{http://www.stsci.edu/hst/wfc3/phot\_zp\_lbn} The WFPC2
magnitudes have been reported to the same photometric system by using
the procedure described in Holtzman et al. (1995), with the gain
settings and zero-points listed in Tab. 28.1 of the HST data handbook.

\section{THE OPTICAL COMPANION TO M28H} 
In order to search for the optical counterpart to the M28H companion,
we carefully re-analyzed a set of $4\arcsec\times 4\arcsec$ WFC3
sub-images centered on the nominal radio position of the MSP.  For
these sub-images the photometric reduction has been re-performed by
using both DAOPHOT (Stetson 1987) and ROMAFOT (Buonanno et al.
1983). In both cases, in order to optimize the identification of faint
objects, we performed the source detection on the median image in the
$U$ band, thus obtaining a master-list. The master-list was then
applied to all the single images in each band and we performed the
PSF-fitting by using appropriate PSF models obtained in each image.
The resulting instrumental magnitudes were reported to those of a
reference image in each filter, and from the frame-to-frame scatter, a
mean magnitude and a standard deviation have been obtained for all the
objects.
  
\begin{figure}[!hp]
\begin{center}
\includegraphics[scale=0.9]{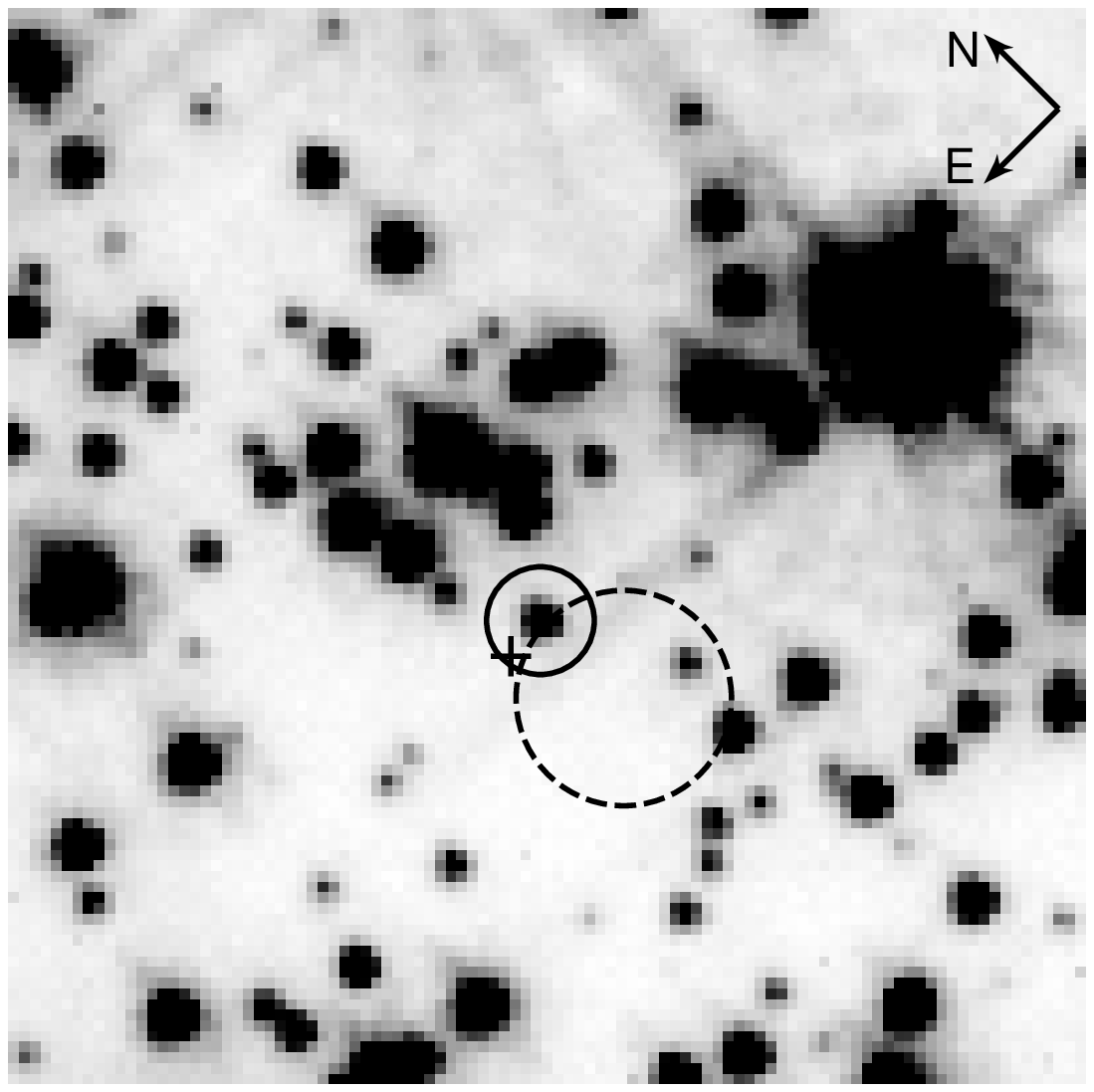}
\caption{$U$-band, $4\arcsec\times4\arcsec$ WFC3 sub-image of M28
  centered on the position of COM-M28H, identified as the
  companion star to the MSP M28H.  The solid circle has a radius of
  $0.2\arcsec$, corresponding to the estimated astrometric accuracy of
  our analysis. The position of the radio source M28H is marked with
  the cross.  The dashed circle ($0.4\arcsec$ radius) marks the
  position and estimated uncertainty of the X-ray source.}
\label{Fig:map}
\end{center}
\end{figure}

We then selected the stars that showed significant variations in the
$U$ band, looking for those that have a periodic variability
compatible with the orbital period of M28H.  Only one object has been
found to match such requirements. This star is located at
$\alpha_{2000}=18^{\rm h} 24^{\rm m} 31.60^{\rm s}$ and
$\delta_{2000}=-24^\circ 52' 17.2\arcsec$, just $0.17\arcsec$ from the
radio position of M28H, and $\sim 0.4\arcsec$ from the X-ray source
(Figure \ref{Fig:map}).

In order to carefully investigate the optical modulation of this star,
we first folded all the measured datapoints obtained in each filter by
using the orbital period $P$ and the reference epoch
($T_0=53755.226397291$ MJD; B\'egin et al. 2010) of the M28H radio
ephemeris.  Since the few measured datapoints in the single bands are
not sufficient to properly cover the entire period (see dots in
Fig. \ref{Fig:lc_filters}), we computed the average magnitude in each
filter and the shift needed to make it match the mean $U$-band
magnitude, which we adopted as a reference. We thus obtained the
combined light curve shown in Figure \ref{Fig:lc_combined}, which well
samples the entire period of variation. Indeed, the optical modulation
of the identified star nicely agrees with the orbital period of the
MSP, thus fully confirming that the variability is associated with the
pulsar binary motion. Hence we conclude that the identified star
(hereafter COM-M28H) is the optical companion to M28H.

This object (with $U=21.99,~V=20.58,~ I=19.49$) is $\sim 1.5$ mag
fainter than the cluster turn-off and slightly bluer than the MS
(Figure \ref{Fig:cmd}). Clearly, it is far too red and bright to be
compatible with a WD (the recycling by-product expected from the
  standard scenario).  Instead, its position in the CMD is marginally
  consistent with a MS star. In particular, the cluster MS is well
  reproduced by a $t=13$ Gyr isochrone (from Marigo et al.  2008),
  with metallicity [Fe/H]$=-1.27$ (from Zinn 1980, after calibration
  to the scale of Carretta \& Gratton 1997, following Ferraro et
  al. 1999) and assuming a color excess $E(B-V)=0.4$ and a distance
  modulus (m-M)$_V=14.97$ (Harris 1996).  By projecting the observed
  magnitudes and colours of COM-M28H onto this isochrone, the
  resulting mass, temperature and radius would be $M_{\rm COM}\sim0.68
  M_\odot$, $T\sim 6000$ K, $R\sim 0.64 R_\odot$, respectively.
  However these quantities should be considered as just an indication,
  since the observational properties of this object (see below)
  strongly suggest that it is a highly perturbed star.

Since the two WFPC2 data-sets have been obtained in two different
epochs with a time baseline of more than 10 years, and given the
relative small distance of M28 (d=5.6 Kpc; Harris 1996), we have been
able to perform a proper-motion analysis.  As shown in Figure
\ref{Fig:proper_motion}, the bulk of stars lie around the position
($\mu_\alpha cos(\delta)=0$, $\mu_\delta=0$) [mas/yr], within a radius
$\sigma_{\mu} \sim1$.  These most likely are members of the cluster,
while field stars are clearly separated (the position in the CMD of
these two classes of stars further confirms such a conclusion;
Dalessandro et al. 2010).  Since no extra-galactic source can be
identified in the FOV adopted for this analysis, no absolute proper
motion determination can be obtained. However a rough estimate can be
derived by averaging the positions of field stars: we obtain
$\mu_\alpha cos(\delta)=-1.40$ and $\mu_\delta=3.50$) [mas/yr], in
agreement with previous results (Cudworth \& Hanson 1993). COM-M28H
lies at ($\mu_\alpha cos(\delta)=-0.09 \pm 0.15$, $\mu_\delta=0.09 \pm
0.15$) [mas/yr], thus fully behaving as a member of the cluster.

\section{DISCUSSION} 
The observed light curve of COM-M28H (Fig. \ref{Fig:lc_combined})
clearly shows two distinct and asymmetric minima, at phases $\phi\sim
0.25$ and $\phi\sim 0.75$, quite similar to what is observed for two
other MSP companions (Ferraro et al. 2001b; Cocozza et al. 2008).
Such a shape is a clear signature of ellipsoidal variations induced by
the NS tidal field on a highly perturbed, bloated star. Moreover, the
relative deepness of the two minima (consistent with a light curve
purely due to ellipsoidal variations) suggests that only a marginal
(if any) over-heating is affecting the side of the companion facing
the pulsar.  This is also supported by the non-detection of $H_\alpha$
emission from the system (see the right panel of Fig. 4).

\begin{figure}[!hp]
\begin{center}
\includegraphics[scale=0.7]{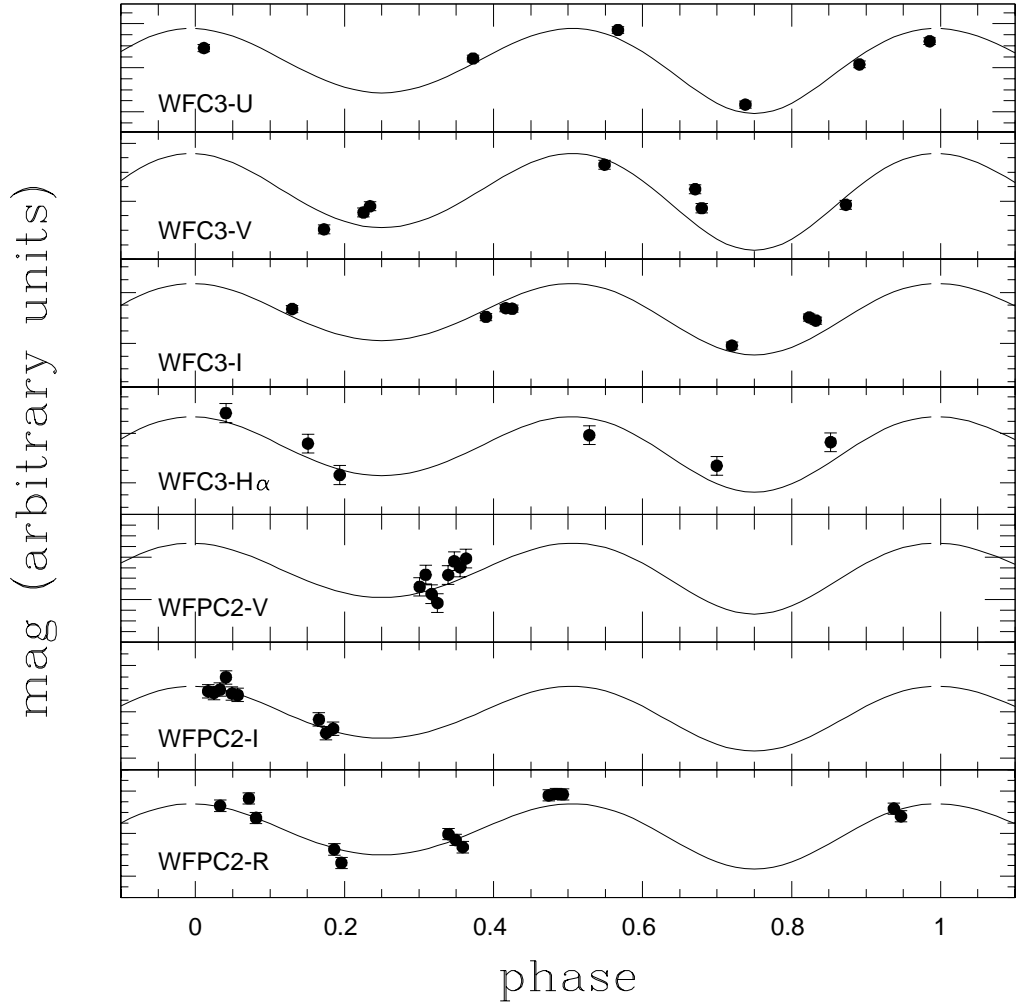}
\caption{The observed light curves in the WFC3 and WFPC2 images. The best
fit model is shown as a solid line in each panel.}
\label{Fig:lc_filters}
\end{center}
\end{figure}

\begin{figure}[!hp]
\begin{center}
\includegraphics[scale=0.7]{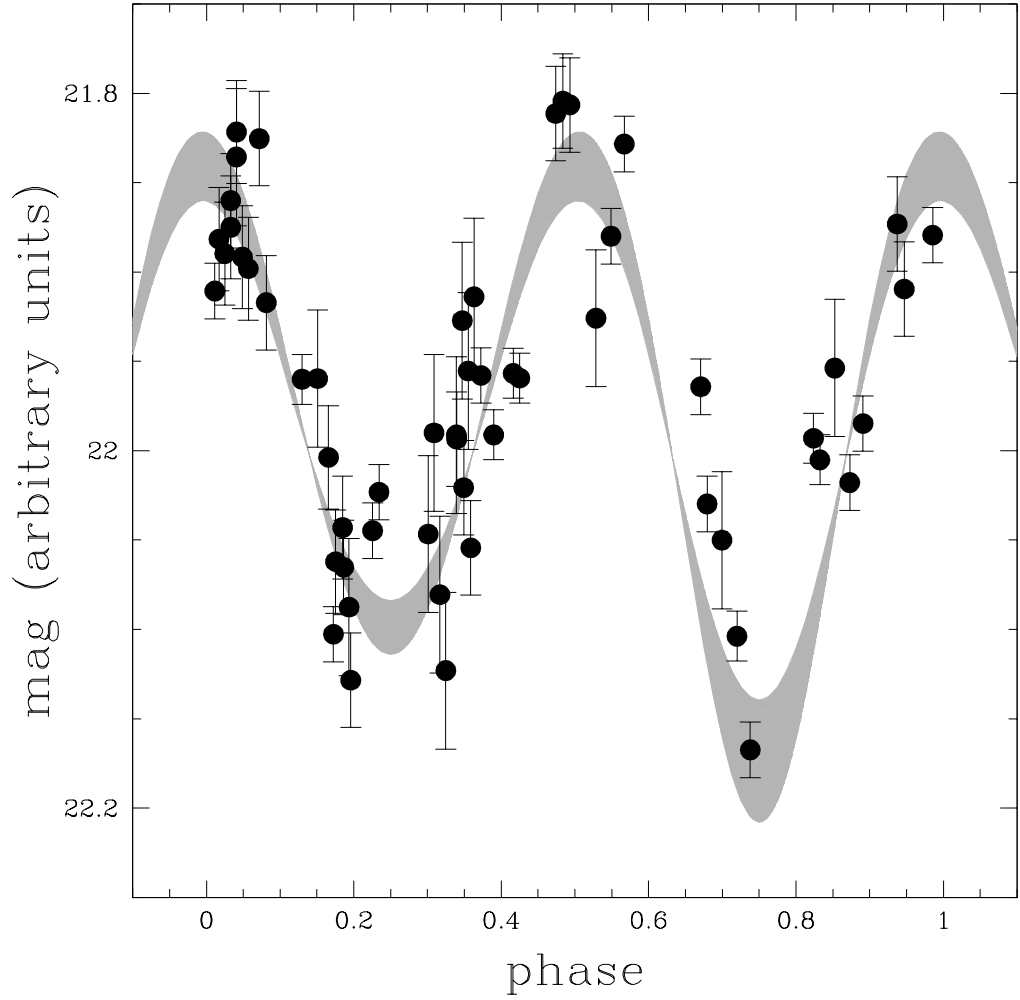}
\caption{The global light curve of COM-M28H, obtained by combining the
  data points shown in Fig. \ref{Fig:lc_filters}. The grey area
  represents the region spanned by the best-fit light curves in each
  photometric band.}
\label{Fig:lc_combined}
\end{center}
\end{figure}

\begin{figure}[!hp]
\begin{center}
\includegraphics[scale=0.7]{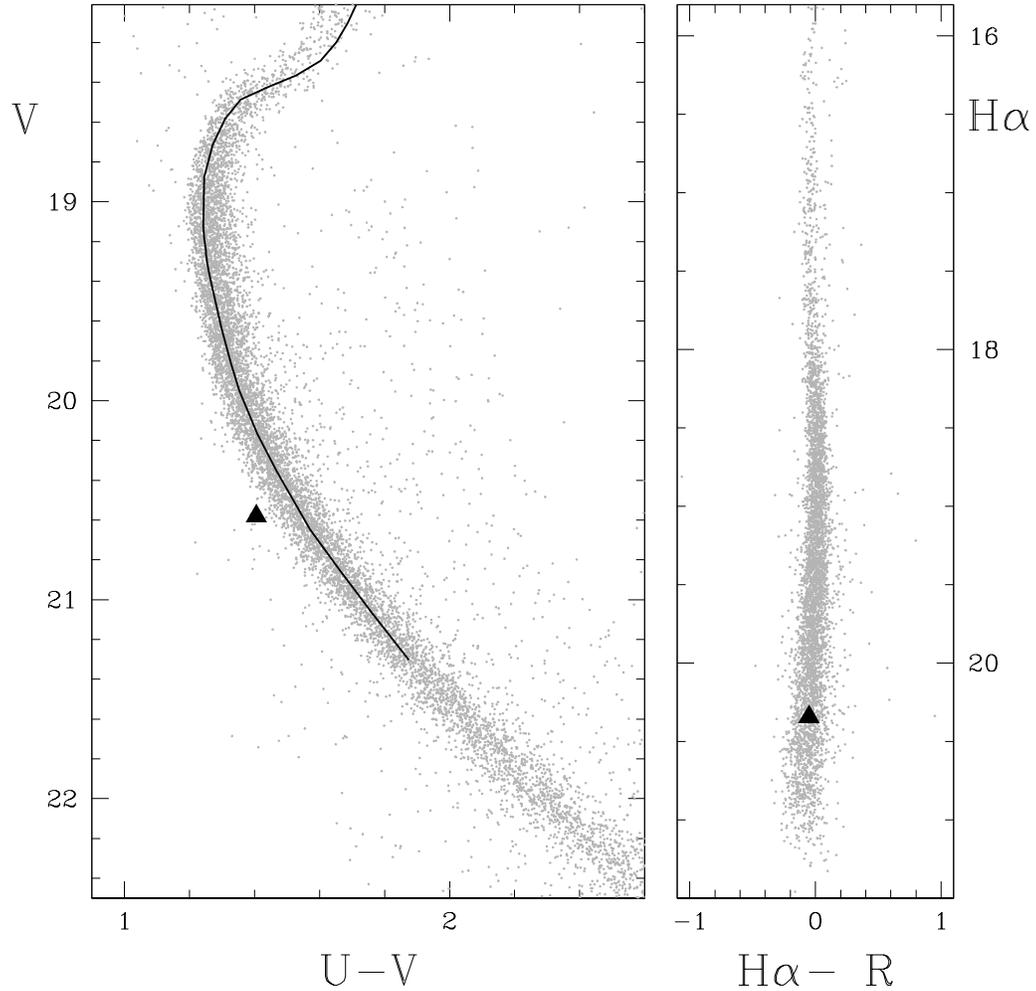}
\caption{CMDs of M28 derived by WFC3 data in a circular region of
  $\sim 30\arcsec$ radius centered on the position of COM-M28H (left
  panel), and by a combination of WFC3 and WFPC2 data (right
  panel). The solid triangle marks the position of COM-M28H.  The
  plotted isochrone (from Marigo et al. 2008) has been obtained for
  $t=13$ Gyr, [Fe/H]$=-1.27$, $E(B-V)=0.4$ and (m-M)$_V=14.97$.  }
\label{Fig:cmd}
\end{center}
\end{figure}

\begin{figure}[!hp]
\begin{center}
\includegraphics[scale=0.7]{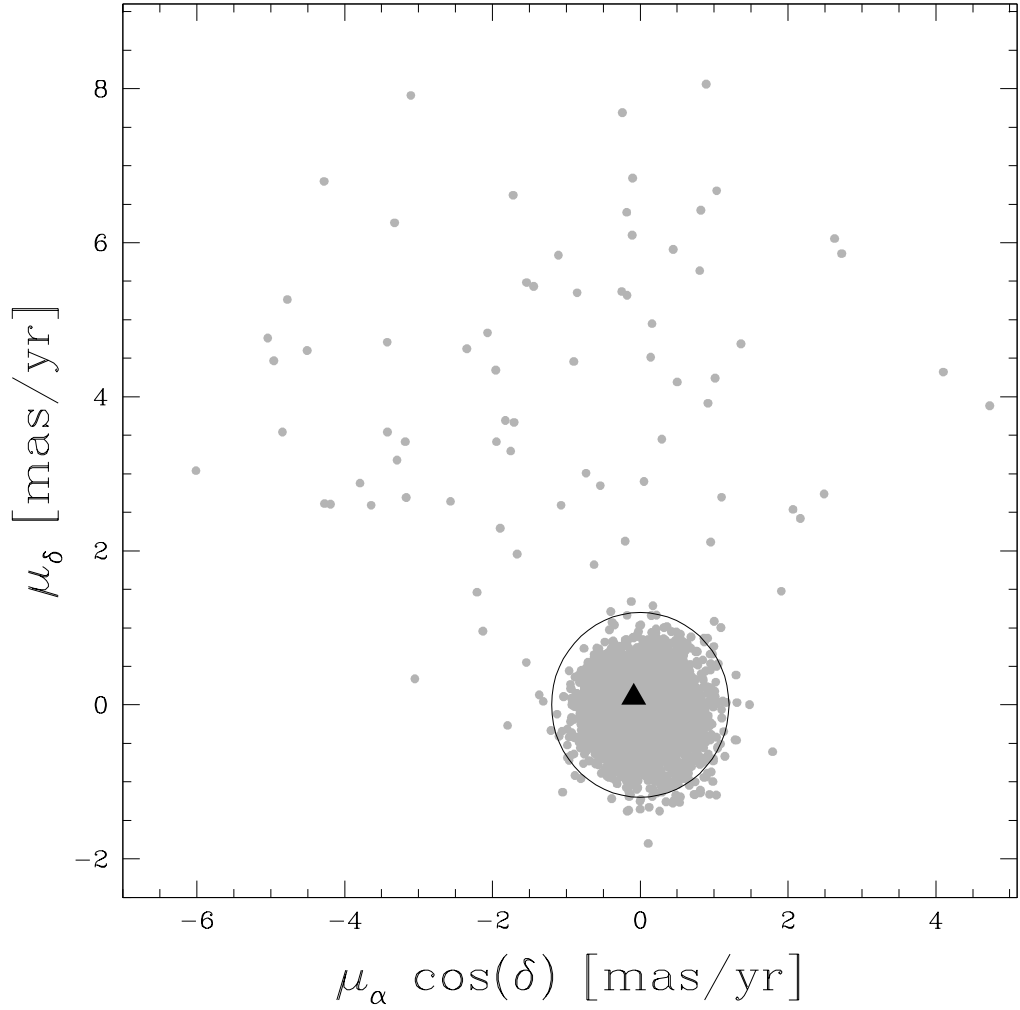}
\caption{Relative proper motion diagram of M28 
in equatorial coordinates. The solid triangle represents the position of COM-M28H. }
\label{Fig:proper_motion}
\end{center}
\end{figure}

Given the mass function derived from the radio observations
($f_1=0.00211277 M_\odot$; B\'egin 2006), and assuming $1.4 M_\odot$
for the MSP mass and $0.68 M_\odot$ for the mass of COM-M28H (as
derived from the cluster best-fit isochrone), the resulting orbital
inclination of the system would be $i\sim 18\arcdeg$.  Such a low
value for the inclination angle would not produce any optical
modulation.  Indeed both the light curve shape and the occurrence of
eclipses in the radio signal point toward a significantly higher value
of the orbital inclination (which corresponds to a lower companion
mass for a given mass function).  By assuming $i=60\arcdeg$ (the
median of all possible inclination angles) a companion mass of $\sim
0.2 M_\odot$ is obtained.  In this configuration the corresponding
total mass of the system would be $M_{\rm T}=1.6 M_\odot$ and the
physical orbital separation of the system is $a\sim 2.8 R_\odot$.  In
order to check whether such a configuration reproduces the observed
light curve, we employed the publicly available software
NIGHTFALL.\footnote{Available at
  http://www.hs.uni-hamburg.de/DE/Ins/Per/Wichmann/Nightfall.html.}.
We fixed the orbital period of the pulsar to the radio value and the
surface temperature of COM-M28H to $T=6000$ K (as inferred from the
position in the CMD). We then used an iterative procedure letting the
orbital inclination, the mass ratio ($M_{\rm NS}/M_{\rm COM}$) and the
Roche Lobe filling factor vary respectively in the ranges
$0\arcdeg-90\arcdeg$, $1-20$, and $0.1-1$.  By using as selection
criterion a $\chi^2$ test, the best-fit model
(Figs. \ref{Fig:lc_filters} and \ref{Fig:lc_combined}) was obtained
for an inclination $i\simeq 65\arcdeg$, a mass ratio $M_{\rm
  NS}/M_{\rm COM}\simeq 7$ and a Roche Lobe filling factor equal to
1. These results confirm that a configuration with a highly distorted
companion of about $0.2 M_\odot$, orbiting a $1.4 M_\odot$ MSP, in a
plane with an orbital inclination of $\sim 60-70\arcdeg$ well
reproduces both the mass function of the system derived from the radio
observations, and the optical light curve of COM-M28H.  It is worth
noticing that a good fit can be obtained only if COM-M28H completely
filled its Roche Lobe, which, following Paczynski (1971), we estimate
to be $\sim 0.65 R_\odot$. While such a large value of the stellar
radius allows to account for the observed luminosity of
COM-28H,\footnote{Under the assumption of black body radiation (with
  the luminosity $L$ given by $L\propto R^2 T_{eff}^4$), a $0.2
  M_\odot$ star heated to the observed temperature ($T_{eff}\sim 6000$
  K) and bloated to a radius of $\sim 0.65 R_\odot$, has a luminosity
  which is fully consistent with the observed one. This reinforces the
  hypothesis that COM-M28H completely filled its Roche Lobe. For sake
  of comparison, the same object with a radius $R\sim 0.2 R_\odot$
  (the value expected for a MS $0.2 M_\odot$; Marigo et al. 2008),
  would be a factor of $\sim 10$ too faint.} it is far too small to
cause the observed radio eclipse. In fact an eclipse lasting for $\sim
20\%$ of the orbital period corresponds to an eclipsing region of
$\sim 3.3 R_\odot$ size. This suggests that the eclipsing material is
extending well beyond the Roche Lobe and that it is probably
constantly replenished (see also B\'egin 2006). Indeed, under the
influence of the MSP intense radiation field, a (otherwise normal) MS
star may expand to fill its Roche Lobe (D'antona \& Ergma 1993) and
even start to lose mass, while the accretion on the pulsar is
inhibited by its magnetic pressure (as in the case of MSP-A in NGC
6397; Ferraro et al 2001b; see also Archibald et al. 2009). Moreover
B\'egin (2006) found a large orbital period derivative for this
system, suggesting that the binary is losing material and is spiraling
out to longer orbital period.

All these considerations indicate that COM-M28H is a
highly-perturbed star which is currently losing mass, and that the
system is surrounded by large clouds of gas. Whether or not part of
the lost mass was accreted by the NS and served to reaccelerate it in
the past (as in the case of J1023+0038; Archibald et al. 2009) cannot
be inferred from the available data. We note however that, while from
the natural cluster dynamical evolution massive objects are expected
to be concentrated close to the centre, M28H is the second most
off-centered (after M28F; B\'egin et al. 2010) and it is located
outside the cluster core. Hence, such an offset position may suggest
the following scenario: the NS was recycled by another companion (that
eventually became a very low mass, exhausted star, because of the
heavy mass transfer); then an exchange interaction occurred in the
cluster core between the MSP binary and a MS star, thus causing the
ejection of the lightest star and kicking the newly-formed system away
from the centre; the new companion started to suffer heavy
perturbations (bloating, mass loss, etc.)  induced by the MSP and we
currently observe it as COM-M28H; then it eventually will become a
helium WD.  If the MSP was ejected from the cluster core in an
exchange encounter, its current position suggests a relatively recent
epoch for the formation of this system, since the expected time for
such a heavy system to sink back into the M28 centre should be lower
than a few Gyr.

While a spectroscopic follow-up may help to better clarify this
scenario, the identification of COM-M28H further increases the number
of MSPs with a non-degenerate companion in GCs: 4 out 7 optical
counterparts identified so far seem to be MS or sub-giant branch,
perturbed stars. This likely indicates that exchange interactions are
common events in the dense environments of GC cores and they are quite
effective in modifying the ``natural" outcomes of the pulsar recycling
processes (Freire 2005).
 
\acknowledgements We thank the anonymous referee for the fast and
careful reading of the manuscript. This research was supported by
Agenzia Spaziale Italiana (under contract ASI-INAF I/009/10/0), by the
Istituto Nazionale di Astrofisica (INAF, under contract PRIN-INAF
2008) and by the Ministero dell'Istruzione, dell'Universit\`a e della
Ricerca. Pulsar research at UBC is supported by an NSERC Discovery
Grant and by the CFI.

\newpage




\begin{thebibliography}{} 

\bibitem[Alpar et al.(1982)]{1982Natur.300..728A} Alpar, M.~A., Cheng, 
A.~F., Ruderman, M.~A., \& Shaham, J.\ 1982, \nat, 300, 728 

\bibitem[Archibald et al.(2009)]{2009Sci...324.1411A} Archibald, A.~M., et 
al.\ 2009, Science, 324, 1411 

\bibitem[Bailyn(1992)]{1992ApJ...392..519B} Bailyn, C.~D.\ 1992, \apj,
  392, 519

\bibitem[Bailyn(1995)]{1995ARA&A..33..133B} Bailyn, C.~D.\ 1995,
  \araa, 33, 133

\bibitem[B\'egin (2006)]{begin 2006} B\'egin, S., M.Sc. thesis,
  Dept. of Physics and Astronomy, University of British Columbia
  (2006).

\bibitem[Bellini \& Bedin(2009)]{2009PASP..121.1419B} Bellini, A., \&
  Bedin, L.~R.\ 2009, \pasp, 121, 1419

\bibitem[Bhattacharya \& van den Heuvel(1991)]{1991PhR...203....1B}
  Bhattacharya, D., \& van den Heuvel, E.~P.~J.\ 1991, \physrep, 203,
  1

\bibitem[Bogdanov et al.(2005)]{2005ApJ...630.1029B} Bogdanov, S.,
  Grindlay, J.~E., \& van den Berg, M.\ 2005, \apj, 630, 1029

\bibitem[Bogdanov et al.(2010)]{bogd10} Bogdanov, S. et al.\ 2010,
  \apj submitted

\bibitem[Buonanno et al. (1983)]{buonanno83} Buonanno, R., Buscema,
  G., Corsi, C.~E, Ferraro, I., \& Iannicola, G., 1983, A\&A, 126, 278

\bibitem[Camilo et al.(2000)]{2000ApJ...535..975C} Camilo, F., Lorimer, 
D.~R., Freire, P., Lyne, A.~G., \& Manchester, R.~N.\ 2000, \apj, 535, 975 

\bibitem[Carretta \& Gratton(1997)]{1997A&AS..121...95C} Carretta, E.,
  \& Gratton, R.~G.\ 1997, \aaps, 121, 95

\bibitem[Clark(1975)]{1975ApJ...199L.143C} Clark, G.~W.\ 1975, \apjl,
  199, L143

\bibitem[Cocozza et al.(2008)]{2008ApJ...679L.105C} Cocozza, G., Ferraro, 
F.~R., Possenti, A., Beccari, G., Lanzoni, B., Ransom, S., Rood, R.~T., 
\& D'Amico, N.\ 2008, \apjl, 679, L105 

\bibitem[Cudworth \& Hanson(1993)]{1993AJ....105..168C} Cudworth,
  K.~M., \& Hanson, R.~B.\ 1993, \aj, 105, 168

\bibitem[D'Antona \& Ergma(1993)]{1993A&A...269..219D} D'Antona, F.,
  \& Ergma, E.\ 1993, \aap, 269, 219
 
\bibitem[Davies \& Hansen(1998)]{1998MNRAS.301...15D} Davies, M.~B.,
  \& Hansen, B.~M.~S.\ 1998, \mnras, 301, 15

\bibitem[Edmonds et al.(2001)]{2001ApJ...557L..57E} Edmonds, P.~D., 
Gilliland, R.~L., Heinke, C.~O., Grindlay, J.~E., 
\& Camilo, F.\ 2001, \apjl, 557, L57  

\bibitem[Edmonds et al.(2002)]{2002ApJ...579..741E} Edmonds, P.~D., 
Gilliland, R.~L., Camilo, F., Heinke, C.~O., 
\& Grindlay, J.~E.\ 2002, \apj, 579, 741 
  
\bibitem[Ferraro et al.(1995)]{1995A&A...294...80F} Ferraro, F.~R.,
  Fusi Pecci, F., \& Bellazzini, M.\ 1995, \aap, 294, 80

\bibitem[Ferraro et al.(1999)]{1999AJ....118.1738F} Ferraro, F.~R.,
  Messineo, M., Fusi Pecci, F., de Palo, M.~A., Straniero, O.,
  Chieffi, A., \& Limongi, M.\ 1999, \aj, 118, 1738
   
\bibitem[Ferraro et al. (2001a)]{ferraro01a} Ferraro, F.~R., D'Amico,
  N., Possenti, A., Mignani, R.P., \& Paltrinieri, B., 2001a, ApJ,
  561, 337
   
\bibitem[Ferraro et al. (2001b)]{ferraro01b} Ferraro, F.~R., Possenti,
  A., D'Amico, N., \& Sabbi, E., 2001b, ApJ, 561, L93

\bibitem[Ferraro et al.(2003a)]{2003ApJ...595..179F} Ferraro, F.~R.,
  Possenti, A., Sabbi, E., Lagani, P., Rood, R.~T., D'Amico, N., \&
  Origlia, L.\ 2003, \apj, 595, 179

\bibitem[Ferraro et al.(2003b)]{2003ApJ...596L.211F} Ferraro, F.~R.,
  Possenti, A., Sabbi, E., \& D'Amico, N.\ 2003, \apjl, 596, L211

\bibitem[Ferraro et al.(2009a)]{2009Natur.462.1028F} Ferraro, F.~R.,
  et al.\ 2009a, \nat, 462, 1028
 
\bibitem[Ferraro et al.(2009b)]{2009Natur.462..483F} Ferraro, F.~R.,
  et al.\ 2009b, \nat, 462, 483
   
\bibitem[Fregeau(2008)]{2008ApJ...673L..25F} Fregeau, J.~M.\ 2008,
  \apjl, 673, L25

\bibitem[Freire et al.(2003)]{2003MNRAS.340.1359F} Freire, P.~C., Camilo, 
F., Kramer, M., Lorimer, D.~R., Lyne, A.~G., Manchester, R.~N., 
\& D'Amico, N.\ 2003, \mnras, 340, 1359 

\bibitem[Freire(2005)]{2005ASPC..328..405F} Freire, P.~C.~C.\ 2005,
  Binary Radio Pulsars, 328, 405

\bibitem[Goodman \& Hut(1989)]{1989Natur.339...40G} Goodman, J., \&
  Hut, P.\ 1989, \nat, 339, 40

\bibitem[Harris(1996)]{harr96} Harris, W.~E. 1996, AJ, 112, 1487

\bibitem[Heggie(1975)]{1975MNRAS.173..729H} Heggie, D.~C.\ 1975,
  \mnras, 173, 729

\bibitem[Hills \& Day(1976)]{1976ApL....17...87H} Hills, J.~G., \&
  Day, C.~A.\ 1976, \aplett, 17, 87

\bibitem[Holtzman et al.(1995)]{1995PASP..107.1065H} Holtzman, J.~A.,
  Burrows, C.~J., Casertano, S., Hester, J.~J., Trauger, J.~T.,
  Watson, A.~M., \& Worthey, G.\ 1995, \pasp, 107, 1065

\bibitem[Hut et al.(1992)]{1992PASP..104..981H} Hut, P., et al.\ 1992,
  \pasp, 104, 981

\bibitem[Ivanova et al.(2008)]{2008MNRAS.386..553I} Ivanova, N.,
  Heinke, C.~O., Rasio, F.~A., Belczynski, K., \& Fregeau,
  J.~M.\ 2008, \mnras, 386, 553

\bibitem[Lanzoni et al.(2007)]{2007ApJ...663.1040L} Lanzoni, B., et al.\ 
2007, \apj, 663, 1040 


\bibitem[Lanzoni et al.(2010)]{2010ApJ...717..653L} Lanzoni, B., et al.\ 
2010, \apj, 717, 653 
 
\bibitem[Lyne et al.(1987)]{1987Natur.328..399L} Lyne, A.~G.,
  Brinklow, A., Middleditch, J., Kulkarni, S.~R., \& Backer,
  D.~C.\ 1987, \nat, 328, 399

\bibitem[Marigo et al.(2008)]{2008A&A...482..883M} Marigo, P.,
  Girardi, L., Bressan, A., Groenewegen, M.~A.~T., Silva, L., \&
  Granato, G.~L.\ 2008, \aap, 482, 883

\bibitem[Meylan \& Heggie(1997)]{1997A&ARv...8....1M} Meylan, G., \&
  Heggie, D.~C.\ 1997, \aapr, 8, 1

\bibitem[Paczy{\'n}ski(1971)]{1971ARA&A...9..183P} Paczy{\'n}ski,
  B.\ 1971, \araa, 9, 183

\bibitem[Phinney(1992)]{1992RSPTA.341...39P} Phinney, E.~S.\ 1992,
  Royal Society of London Philosophical Transactions Series A, 341, 39


\bibitem[Pooley et al.(2003)]{2003ApJ...591L.131P} Pooley, D., et
  al.\ 2003, \apjl, 591, L131

\bibitem[Possenti et al.(2003)]{2003ApJ...599..475P} Possenti, A.,
  D'Amico, N., Manchester, R.~N., Camilo, F., Lyne, A.~G., Sarkissian,
  J., \& Corongiu, A.\ 2003, \apj, 599, 475

\bibitem[Pryor \& Meylan(1993)]{1993ASPC...50..357P} Pryor, C., \&
  Meylan, G.\ 1993, Structure and Dynamics of Globular Clusters, 50,
  357

\bibitem[Ransom et al.(2005)]{2005Sci...307..892R} Ransom, S.~M.,
  Hessels, J.~W.~T., Stairs, I.~H., Freire, P.~C.~C., Camilo, F.,
  Kaspi, V.~M., \& Kaplan, D.~L.\ 2005, Science, 307, 892
 
\bibitem[Sigurdsson et al.(2003)]{2003Sci...301..193S} Sigurdsson, S.,
  Richer, H.~B., Hansen, B.~M., Stairs, I.~H., \& Thorsett,
  S.~E.\ 2003, Science, 301, 193

\bibitem[Stairs et al.(2006)]{2006AAS...20915902S} Stairs, I.~H.,
  B\'egin, S., Ransom, S., Freire, P., Hessels, J., Katz, J., Kaspi,
  V., \& Camilo, F.\ 2006, Bulletin of the American Astronomical
  Society, 38, 1118

\bibitem[Stetson(1987)]{1987PASP...99..191S} Stetson, P.~B.\ 1987,
  \pasp, 99, 191

\bibitem[Stetson(1994)]{Stet1994} Stetson, P.~B.\ 1994, \pasp, 106,
  250
 
\bibitem[Verbunt et al.(1997)]{1997A&A...327..602V} Verbunt, F., Bunk,
  W.~H., Ritter, H., \& Pfeffermann, E.\ 1997, \aap, 327, 602

\bibitem[Zinn(1980)]{1980ApJS...42...19Z} Zinn, R.\ 1980, \apjs, 42, 19 

\end{thebibliography}
\end{document}